\documentclass[11pt,a4paper]{article}
\usepackage{jcappub}

\usepackage{amsmath}
\usepackage{amssymb}
\usepackage{amscd}
\usepackage{mathrsfs}
\usepackage{graphicx}
\usepackage{subfigure}
\usepackage{slashed}
\usepackage{mathbbol}
\usepackage{subfigure}
\usepackage{enumerate}
\usepackage{multirow}

\def\aG{ \alpha_{_G} }
\def\gG{ g_{_G}  }
\def\MG{ M_{_G}  }
\def\aD{ \alpha_{_D} } 
\def\mD{ \m_{_D} } 
\def\HD{ H_{_D}  } 
\def\gD{ \g_{_D} } 
\def\aBL{\alpha_{_{B-L}}  }
\def\gBL{ g_{_{B-L}}  }
\def\MBL{ M_{_{B-L}}  }
\def\ZBL{ Z_{_{B-L}}' }

\def\Bv{B_1}
\def\Lv{L_1}
\def\BLv{(B-L)_1}
\def\Bd{B_2}

\def\gv{g_1}
\def\gd{g_2}
\def\gvd{g_{1,\rm{dec}}}
\def\gdd{g_{2,\rm{dec}}}
\def\gvN{g_{1,\rm{BBN}}}
\def\gdN{g_{2,\rm{BBN}}}
\def\Tdec{T_\rm{dec}}
\def\Tv{T_1}
\def\Td{T_2}

\def\yv{y_1}
\def\yd{y_2}

\title{Visible and dark matter from a first-order phase transition in a baryon-symmetric universe}

\author[1]{Kalliopi Petraki,}
\author[2]{Mark Trodden}
\author[1]{and Raymond R. Volkas}

\affiliation[1]{ARC Centre of Excellence for Particle Physics at the Terascale, School of Physics, The University of Melbourne, Victoria 3010, Australia}
\affiliation[2]{Center for Particle Cosmology,  Department of Physics and Astronomy, University of Pennsylvania, Philadelphia, Pennsylvania 19104, USA}

\emailAdd{kpetraki@unimelb.edu.au}
\emailAdd{trodden@physics.upenn.edu}
\emailAdd{raymondv@unimelb.edu.au}

\date{\today}

\abstract{
The similar cosmological abundances observed for visible and dark matter suggest a common origin for both. By viewing the dark matter density as a dark-sector asymmetry, mirroring the situation in the visible sector, we show that the visible and dark matter asymmetries may have arisen simultaneously through a first-order phase transition in the early universe. The dark asymmetry can then be equal and opposite to the usual visible matter asymmetry, leading to a universe that is symmetric with respect to a generalised baryon number.  We present both a general structure, and a precisely defined example of a viable model of this type.  In that example, the dark matter is ``atomic'' as well as asymmetric, and various cosmological and astrophysical constraints are derived.  Testable consequences for colliders include a $Z'$ boson that couples through the $B-L$ charge to the visible sector, but also decays invisibly to dark sector particles.  The additional scalar particles in the theory can mix with the standard Higgs boson and provide other striking signatures.  
}

\begin{document}
\maketitle
\newcommand{\nc}{\newcommand}
\def\cal{\mathcal}
\def\rm{\mathrm}

\nc{\eq}[1]{Eq.~(\ref{#1})}
\nc{\eqs}[1]{Eqs.~(\ref{#1})}

\nc{\pd}{\partial}
\nc{\bea}{\begin{eqnarray}}
\nc{\eea}{\end{eqnarray}}
\nc{\bal}{\begin{alignedat}}
\nc{\eal}{\end{alignedat}}
\nc{\beq}{\begin{equation}}
\nc{\eeq}{\end{equation}}
\nc{\bit}{\begin{itemize}}
\nc{\eit}{\end{itemize}}
\nc{\benu}{\begin{enumerate}}
\nc{\eenu}{\end{enumerate}}
\nc{\bdes}{\begin{description}}
\nc{\edes}{\end{description}}

\nc{\nn}{\nonumber}

\nc{\hc}{\rm{h.c.}}
\nc{\cc}{\rm{c.c.}}

\nc{\sub}[1]{_{\rm{#1}}}
\nc{\ssub}[1]{_{_\rm{#1}}}
\nc{\super}[1]{^{\rm{#1}}}
\nc{\ssuper}[1]{^{^\rm{#1}}}


\nc{\pare}[1]{\left( #1 \right)}
\nc{\sqpare}[1]{\left[ #1 \right]}
\nc{\ang}[1]{\langle #1 \rangle}
\nc{\abs}[1]{\left| #1 \right|}

\def\g5{\gamma_{5}}

\def \eV{\: \rm{eV}}
\def\keV{\: \rm{keV}}
\def\MeV{\: \rm{MeV}}
\def\GeV{\: \rm{GeV}}
\def\TeV{\: \rm{TeV}}

\def\erg{\: \rm{erg}}

\def \cm{\: \rm{cm}}
\def \km{\: \rm{km}}
\def \pc{\: \rm{pc}}
\def\kpc{\: \rm{kpc}}
\def\Mpc{\: \rm{Mpc}}
\def\Gpc{\: \rm{Gpc}}
\def \AU{\: \rm{A.U.}}

\def\sr{\: \rm{sr}}

\def \snd{\: \rm{s}}
\def  \yr{\: \rm{yr}}
\def \Myr{\: \rm{Myr}}
\def \Gyr{\: \rm{Gyr}}

\def \gr {\: \rm{g}}
\def \kgr{\: \rm{kg}}

\def\a{\alpha}
\def\b{\beta}
\def\g{\gamma}
\def\d{\delta}
\def\e{\epsilon}
\def\z{\zeta}
\def\h{\eta}
\def\th{\theta}
\def\i{\iota}
\def\k{\kappa}
\def\l{\lambda}
\def\m{\mu}
\def\n{\nu}
\def\ks{\xi}
\def\om{o}
\def\p{\pi}
\def\r{\rho}
\def\s{\sigma}
\def\t{\tau}
\def\y{\upsilon}
\def\f{\phi}
\def\x{\chi}
\def\ps{\psi}
\def\w{\omega}

\def\ve{\varepsilon}
\def\vr{\varrho}
\def\vs{\varsigma}
\def\vf{\varphi}

\def\G{\Gamma}
\def\D{\Delta}
\def\Th{\Theta}
\def\L{\Lambda}
\def\Ks{\Xi}
\def\P{\Pi}
\def\S{\Sigma}
\def\Y{\Upsilon}
\def\F{\Phi}
\def\Ps{\Psi}
\def\W{\Omega}


\section{Introduction}

The dark matter (DM) problem is a central one of fundamental physics, pointing unequivocally to the existence of new physics beyond the current paradigm of the standard model (SM) of particle physics plus general relativity. We suppose in this paper that the DM problem points to the existence of new stable particles rather than modified gravity, and explore the hypothesis that the similar observed present-day densities for visible matter (VM) and DM suggest a common explanation for both.  This 
general idea has a history stretching back many years, but has become a particularly active area of investigation in recent years~\cite{Dodelson:1989ii,Dodelson:1989cq,Dodelson:1990ge,Kuzmin:1996he,Kitano:2004sv,Kitano:2005ge,Gu:2007cw,Gu:2009yy,An:2009vq,Davoudiasl:2010am,Gu:2010ft,Heckman:2011sw,Oaknin:2003uv,Farrar:2005zd,Bell:2011tn,Cheung:2011if,vonHarling:2012yn,MarchRussell:2011fi,Kamada:2012ht,Nussinov:1985xr,Barr:1990ca,Barr:1991qn,Dodelson:1991iv,Kaplan:1991ah,Thomas:1995ze,Kusenko:1997si,Enqvist:1998en,Berezhiani:2000gw,Fujii:2002kr,Fujii:2002aj,Foot:2003jt,Foot:2004pq,Hooper:2004dc,Agashe:2004bm,Cosme:2005sb,Suematsu:2005kp,Suematsu:2005zc,Gudnason:2006ug,Gudnason:2006yj,Roszkowski:2006kw,Banks:2006xr,Dutta:2006pt,McDonald:2006if,Berezhiani:2008zza,Khlopov:2008ty,Kitano:2008tk,Ryttov:2008xe,Foadi:2008qv,Kaplan:2009ag,Kribs:2009fy,Cohen:2009fz,Shoemaker:2009kg,Cai:2009ia,Frandsen:2009mi,Gu:2010yf,Dulaney:2010dj,Cohen:2010kn,Shelton:2010ta,Haba:2010bm,Buckley:2010ui,Chun:2010hz,Blennow:2010qp,McDonald:2011zz,Hall:2010jx,Allahverdi:2010rh,Dutta:2010va,Higashi:2011qq,Falkowski:2011xh,Doddato:2011fz,Haba:2011uz,Chun:2011cc,Kang:2011wb,Graesser:2011wi,Frandsen:2011kt,Kaplan:2011yj,Cui:2011qe,Mazumdar:2011zd,Kasuya:2011ix,Kumar:2011np,Graesser:2011vj,Oliveira:2011gn,Arina:2011cu,McDonald:2011sv,Kane:2011ih,Barr:2011cz,Lewis:2011zb,Cui:2011wk,Doddato:2011hx,Cui:2011ab,D'Eramo:2011ec,Kang:2011ny}.

An attractive idea for the origin of the DM density is the so-called ``WIMP coincidence'', where WIMP stands for ``weakly interacting massive particle'', meaning that its self-annihilation cross-section is of roughly weak scale, leading to a thermal relic abundance of the correct magnitude.  This idea has the interesting property that the abundance of dark matter is tied to the idea of new physics at the weak scale, required independently for particle physics reasons. However, in this picture we must find an independent and coincidental origin for the similar abundance of visible matter, generating it through a typically unrelated baryogenesis mechanism in the early universe (see, for example~\cite{Riotto:1999yt}).

An interesting alternative to the WIMP hypothesis is that the present-day DM density is also due to an asymmetry, where the DM is a stable member of a hidden sector, and where the two asymmetries are related. This proposal is often referred to as ``asymmetric dark matter'' (ADM). There are three obvious possibilities for how a relation between the visible and the dark asymmetries could arise:  (i) the VM asymmetry is generated first, with the DM asymmetry arising from a subsequent reprocessing which leaks a portion of the visible charge into the dark sector, (ii) vice-versa, or (iii) both asymmetries arise simultaneously from a common origin.  

In the first two possibilities, the VM and DM final asymmetries add up to the original asymmetry generated in one of the sectors at high energies, with the distribution of charge determined by chemical equilibrium between the two sectors at an intermediate scale. 
In the third case, the VM and DM simultaneously develop asymmetries which compensate each other under a generalised conserved particle number that is common to both VM and DM, so that the relationship between the two asymmetries is completely understood and extremely tight.  
It is then appropriate to talk about a ``baryon-symmetric universe'', where by ``baryon number'' we mean precisely this always conserved generalised particle number. 
The baryonic and antibaryonic charges are \emph{separated} into the visible and the dark sectors respectively through non-equilibrium dynamics in the early universe. In all of the above cases, the transfer operators responsible for the generation, sharing and/or separation of charge are decoupled in the low-energy late universe, and the asymmetries of each sector are thus conserved separately. 

In this work we explore this possibility of a baryon-symmetric universe, first introduced in Refs.~\cite{Dodelson:1989ii,Dodelson:1989cq,Dodelson:1990ge}. 
Previous papers have constructed examples of such scenarios using 
out-of-equilibrium heavy particle decays~\cite{Kuzmin:1996he,Kitano:2004sv,Kitano:2005ge,Gu:2007cw,Gu:2009yy,An:2009vq,Davoudiasl:2010am,Gu:2010ft,Heckman:2011sw} with such scenarios termed ``hylogenesis''~\cite{Davoudiasl:2010am}; the QCD phase transition~\cite{Oaknin:2003uv}; asymmetric freeze-in~\cite{Farrar:2005zd}; 
the Affleck-Dine mechanism~\cite{Bell:2011tn,Cheung:2011if,vonHarling:2012yn} -- with those scenarios termed ``pangenesis''~\cite{Bell:2011tn} -- and a time-dependent background of a light scalar field~\cite{MarchRussell:2011fi,Kamada:2012ht}.
Here, we show that one of the generic ways to produce a particle number asymmetry -- bubble nucleation due to a first-order phase transition -- can also be used in the baryon-symmetric universe context.

We emphasise that while all ADM models relate the visible and dark asymmetries, the dynamics involved is quite different from model to model. This is important, not only from a model-building perspective, but also because of the different observational signatures. The most generic property of baryon-symmetric models is that there is an \emph{always} conserved particle-number symmetry. Because it is always conserved, it is natural that it be a gauge symmetry. The corresponding gauge boson can then potentially be discovered in colliders and identified by its invisible decay into dark sector particles. We discuss the importance of this and other probes further in section~\ref{sec:signatures}.
Other significant signatures of baryon-symmetric models -- albeit more model-dependent -- include baryon destruction in the core of compact objects, and induced nucleon decay in DM direct-detection experiments~\cite{Davoudiasl:2011fj}. In view of the above, it is important to distinguish between the scenario we present here, and that of hidden sector baryogenesis via bubble nucleation explored in Refs.~\cite{Dutta:2006pt,Shelton:2010ta,Dutta:2010va}, in which there is no unbroken particle number carried by both VM and DM.

To connect the VM and DM \emph{mass} densities, two ingredients are needed.  The first, as discussed above, is a close connection between the two asymmetries, and the second is a theory of the DM mass scale.  It is sensible to first investigate generic ways of connecting the asymmetries, and then to think through different hidden sector possibilities for the DM.  Our focus here is thus be on connecting the asymmetries.  For the purposes of illustration, we present an example of a fully-specified dark sector, but one must recognise that a plethora of dark sectors are \emph{a priori} possible.  We do not investigate the deep problem of relating the DM mass scale to VM mass scales in this work.  As for most other asymmetric DM models, our scheme motivates the existence of a GeV-scale DM mass, with some encouragement provided by results from the DAMA, CoGeNT and CRESST direct detection experiments.

In the next section we explain the very simple symmetry structure underpinning our model and indeed all models of a baryon-symmetric universe~\cite{Bell:2011tn}.  Section~\ref{sec:mechanism} then details our model. In section~\ref{sec:constraints} we discuss cosmological and astrophysical constraints pertinent to most models which introduce a separate dark sector. We discuss observational signatures in section~\ref{sec:signatures}, and finish with some concluding remarks.

\section{Symmetry structure for a baryon-symmetric universe}
\label{sec:symmetrystructure}

The visible-matter content of our present universe is understood in terms of the symmetries of the SM. In the SM, baryon number is conserved up to anomalous interactions that are irrelevant at zero temperature. This is the main reason for the stability of ordinary matter, ensuring the stability, or near-stability, of the proton, the lightest particle carrying baryon number.  
Baryon number in the SM is additively conserved, and thus stabilises the proton both against decay and against self-annihilation. 
The SM possesses two other additively conserved Abelian charges, the electric charge and the lepton number, which are responsible for the stability of the electron and the lightest neutrino\footnote{This is the case if neutrinos are Dirac fermions. If neutrinos are Majorana, lepton number is violated, but the lightest neutrino is still stable due to the multiplicatively conserved fermion number. The latter arises from angular momentum conservation.}  respectively.
However, most of the mass density of the VM today is due to protons and heavier nuclei.

Let us call the baryon number of the ordinary sector $\Bv$.  
We now make the hypothesis that the stability of the DM particle is due to an analogous U(1) symmetry, sustaining a conserved, or approximately conserved, ``dark baryon number'', which we call $\Bd$.  Additional stable particles may exist in the dark sector, due to extra U(1) symmetries, in analogy to the electron and the lightest neutrino.

If we are to ever understand the baryon asymmetry of the universe, we must suppose that $\Bv$ conservation did not hold for some period during the early universe (and it may not hold today, but its zero temperature, low-energy violation must be very weak indeed).  Similarly, asymmetric DM requires that $\Bd$ was once violated under the appropriate conditions for the creation of a dark-sector asymmetry.

Now consider the linear combinations
\begin{equation}
B \equiv \Bv - \Bd \quad \rm{and}\quad X \equiv \Bv + \Bd \ .
\label{eq:BandX}
\end{equation}
Suppose that $B$ is strictly conserved during the epoch
of asymmetry generation, and defines a generalised particle number encompassing both the ordinary and the dark sectors that we simply call ``baryon number''.  By ``baryon symmetric universe'' we mean ``$B$ symmetric''.  In this case, we require the violation only of the orthogonal linear combination, $X$, alone in the early universe, and its restoration today.  The Sakharov conditions can be used to guide the construction of mechanisms for generating a nonzero $X$ asymmetry, while maintaining the $B$ number of the universe as zero.  If so, we get an automatic \emph{separation} of baryon number between the ordinary and dark sectors,
\begin{equation}
\Delta \Bv = \Delta \Bd = \frac{\Delta X}{2},\qquad \Delta B = 0 \ .
\label{eq:Bv-Bd separation}
\end{equation}
In this way we have recast the baryon-asymmetry problem as the $X$-asymmetry problem.  
The restoration of $X$ in the low-energy late universe, together with the conservation of $B$, ensures the stability of both VM and DM.

The discussion above took $\Bv$ to be the ordinary baryon number, the symmetry that stabilizes the proton.  But, in fact, $\Bv$ could be any linear combination of the ordinary baryon and lepton numbers, except for their sum. The extreme case is when $\Bv$ is actually lepton number, which means that $X$ asymmetry generation is then leptogenesis as far as the visible sector is concerned.  This is acceptable, of course, because electroweak sphalerons will process the lepton asymmetry partly into an ordinary baryon asymmetry, and the same will hold for all other linear combinations (barring the sum, which would be washed out).  These observations are potentially very important, because the deep reason for the conservation of what we called $B$ in Eq.~(\ref{eq:BandX}) may well be that it is anomaly-free and gauged, in which case $\Bv$ might be better thought of as ordinary baryon-minus-lepton number $\BLv$.
In the explicit model we present in the next section, we shall indeed suppose that ``$B$'' is gauged, and we shall rename it $B-L$ to honour that fact.

Of course, no massless gauge boson that couples to the visible sector as $\BLv$ has been observed. This compels the breaking of the local $B-L$. We require that this breaking leaves a remnant global symmetry unbroken, which for all of the visible and dark sector fields is identical to $B-L = \BLv - \Bd$.  We discuss the breaking of the gauged $B-L$ in section~\ref{sec:B-L}.

\section{The mechanism and its cosmological history}
\label{sec:mechanism}
We now introduce the dynamics and the particle content of our mechanism, and describe its cosmological history. 

In the electroweak baryogenesis scenario, the electroweak phase transition is first-order, proceeding via the nucleation of bubbles of
the electroweak-symmetry-breaking vacuum, which eventually coalesce to produce an approximately homogeneous universe in the broken phase.  During the transition, baryon-number violation is occurring rapidly in the unbroken phase regions due to the sphaleron effect, while it is ineffective in the broken-phase bubbles.  The moving bubble walls lead to departures from thermal equilibrium, and SM fermions interact with the bubble walls via C and CP violating Yukawa interactions. The Sakharov conditions are met, and a baryon asymmetry is generated.

We want to use this kind of dynamical setup to generate an $X$ asymmetry without generating a $B$ asymmetry, hereinafter renamed ``$B-L$'' and assumed to be gauged.  We therefore need a sector which features a non-Abelian gauge force with respect to which $X$ is anomalous, but $B-L$ is not. The sphalerons of this new gauge force, which we call the ``generative interaction'', will then mediate $X$-violating processes, and we arrange for the generative interactions to be spontaneously broken during a first-order phase transition prior to the second-order electroweak phase transition.\footnote{In the SM, the electroweak phase transition is known to be second order, so new physics would need to be introduced to make it first order. Although a first order electroweak phase transition would not change X, it could (in the presence of sufficient C and CP violation) create $\Bv$ and $\Lv$ in equal amounts, thus spoiling the relation between $\Bv$ and $\Bd$ we shall derive in the following sections. We thus assume a second order electroweak phase transition as is predicted in the SM.}

For the purposes of dark and visible matter genesis, we may discern three sectors in our theory:
\benu[(i)]
\item the ordinary or visible sector, which contains particles with $B-L = X$ or $\Bd=0$,  
\item the dark sector, which contains particles with $B-L = -X$ or $\BLv = 0$, and 
\item the generative sector, which contains particles charged under both $\BLv$ and $\Bd$.  
\eenu
We arrange for the $X$ asymmetry produced in the generative sector to cascade down to the lightest visible and dark sector particles via $X$ and $B-L$ conserving interactions of the form: generative $\to$ visible $+$ dark, resulting in equal and opposite visible and dark sector asymmetries.

The particle content of our model consists of the SM plus the new fermions and scalars in Table~\ref{tab:charges} and the associated new gauge bosons. 
Most of this table defines the generative and dark sectors, with the fermions $f_{L,R}$ being additions to the visible sector.  
The SM sector is understood to contain the physics of neutrino mass generation, which we leave unspecified because there are many possibilities. However, we assume the existence of three right-handed neutrinos to cancel the cubic $B-L$ anomalies and gauge $B-L$.
The dark sector content, including the dark U(1)$_D$ interaction, is a minimal choice for a viable asymmetric DM scenario, as will become clear in section~\ref{sec:dark sect}.
There must also be dynamics for breaking the gauged U(1)$_{B-L}$, leaving a remnant stabilising global symmetry, 
as mentioned in section~\ref{sec:symmetrystructure}. We discuss the $B-L$ breaking in section~\ref{sec:B-L}

The interactions and the degrees of freedom of Table~\ref{tab:charges} represent minimal choices, sufficient to illustrate the mechanism and provide a viable dark sector. More elaborate structures can, of course, be built and provide elegant solutions to various secondary issues that will arise during our discussion.

\begin{table}[t]
\begin{center}
\begin{tabular}{|c|l|c|c|c|c|c|}
\hline 
\multirow{2}{*}{Sectors} & \multirow{2}{*}{Particles} & SU(2)$_G$ & U(1)$_{B-L}$ & U(1)$_X$  & U(1)$_D$ & Yukawa  \\
                         &                            & gauged    & gauged       & anomalous & gauged   & interactions\\  
\hline \hline
                         & $\ps_L$                    &  2        &  0           & -2        & 0  & \multirow{2}{*}{$\bar{\ps}_L \, \vf \, \ps_R   +  \bar{\ps}_L \, \tilde{\vf} \, \ps_R $} \\
generative               & $\ps_{1R}, \ps_{2R}$       &  1        &  0           & -2        & 0  &  \\ 
                         & $\vf$                      &  2        &  0           &  0        & 0  & $\bar{\ps}_L \, \x \, f_R$ \\ \hline
\multirow{2}{*}{visible} & $ f_{L,R}$                 &  1        & -1           & -1        & 0  & \multirow{2}{*}{$\bar{\ell}_L \, H \, f_R$} \\ 
                         & $ \n_R$                    &  1        & -1           & -1        & 0  &  \\ \hline
                         & $\x$                       &  2        &  1           & -1        & 0  &  \\ 
dark                     & $\ks_{L,R}$                &  2        &  0           &  0        & 1  & $\bar{\ks} \, \x \, \z$ \\ 
                         & $\zeta_{L,R}$              &  1        & -1           &  1        & 1  &  \\ 
\hline
\end{tabular}
\end{center}
\caption{The charge assignments under the new symmetries in our model. The model consists of the SM fields, the fields in this table and a scalar $\sigma$ that breaks U(1)$_{B-L}$. Three right-handed neutrinos are introduced to cancel the cubic $B-L$ anomaly, and an even number of families of the $\psi$ fermions is required to cancel the Witten anomaly.
The generative-sector field $\varphi$ gains a nonzero VEV during a first-order phase transition, while the dark-sector scalar $\x$ retains always a zero VEV. 
The right column summarises the interactions responsible for the generation and transfer of the $X$ asymmetry.
}
\label{tab:charges}
\end{table}

\subsection{The generative sector}
\label{sec:gen sect}
The generative sector has the gauge symmetry SU(2)$_G$ and the global symmetry U(1)$_X$. Its matter content consists of the chiral fermions $\ps$ Yukawa-coupled to the scalar multiplet $\vf$, as per
\begin{equation}
\d\cal{L}_\rm{G}  
= -\sqpare{ \sum_{j=1}^2 \pare{h_j \bar{\ps}_L \: \vf \: \ps_{jR}  + \tilde{h}_j \bar{\ps}_L \: \tilde{\vf} \: \ps_{jR}} + \hc } 
- V_\rm{G}(\vf) \ , 
\label{eq:LG}
\end{equation}
where $\tilde{\vf} \equiv i\tau_2 \vf^*$, and $V_\rm{G}(\vf)$ is a scalar potential for $\vf$.
There need to be at least two families of the $\ps$ fermions for there to be CP-violating phases in these Yukawa interactions, and an even number of families is required to cancel the SU(2)$_G$ Witten anomaly~\cite{Witten:1982fp}.  For $N$ families, the four Yukawa matrices in Eq.~(\ref{eq:LG}) contain a total of $4N^2$ phases.  The rephasings of $\ps_L$, $\ps_{1R}$, $\ps_{2R}$ and $\vf$ will remove $3N+1$ phases, leaving $(4N+1)(N-1)$ physical CP-violating phases.  The minimal choice $N=2$ is free of the Witten anomaly and has nine physical CP-violating phases.

None of the fields in the generative sector carry $B-L$ or $D$.  The global symmetry U(1)$_X$, which is the fermion number of the generative sector\footnote{\label{foot:Xsym} Note that the gauge symmetries and the number of families allow the mass terms $\overline{\ps_L^c} \ps_L^{}, \overline{\ps_R^c} \ps_R^{}$, which violate U(1)$_X$. We impose U(1)$_X$ and eliminate these terms. Although in this model U(1)$_X$ is not accidental, a more elaborate symmetry structure could give rise to an accidental, and anomalous, U(1)$_X$.}, 
has an $[\, \rm{SU}(2)_G\, ]^2 \rm{U}(1)_X$ anomaly, as required. Sphalerons of SU(2)$_G$ mediate rapid $X$-violating processes before the SU(2)$_G$ phase transition takes place, driven by the spontaneous-symmetry-breaking dynamics of $\vf$ that completely break the generative SU(2).

We arrange the parameters of the Higgs potential for $\vf$ to produce a first-order phase transition in the generative sector
at a critical temperature $T_c \sim \langle \vf \rangle \equiv v_\vf$, which is at the TeV scale or above.  
Bubbles of broken SU(2)$_G$ phase are nucleated, and we are free to choose the parameters so that the phase transition is very strongly first order.  In that case, the rapid SU(2)$_G$ sphaleron-mediated $X$-violating transitions in the remaining unbroken-phase
regions, which occur at the rate~\cite{Moore:1997sn}
\beq
\G_\rm{G \, sphaleron} \simeq 30\, \aG^5\, T^4 \ ,
\eeq
get switched off as soon as a bubble-wall passes.  The parameter $\aG$ is the generative fine-structure constant
$\aG \equiv \gG^2/4\pi$, where $\gG$ is the gauge coupling constant.

The $\ps$ fermions interact with the moving bubble walls through the CP-violating Yukawa interactions of Eq.~(\ref{eq:LG}), creating an $X$ charge carried by those fermions.  The competing washout process is very ineffective if the transition is strongly first order.  By choosing the CP-violating phases appropriately, we are able to generate an $X$ asymmetry of the required size,
\beq 
\h(X) \equiv \frac{n_X - n_{\bar{X}}}{s} \sim 10^{-9} \ ,
\eeq
where $s$ is entropy density of the universe. That this is possible is clear from the extensive literature on electroweak baryogenesis~\cite{Riotto:1999yt}, the dynamics of which we are mimicking here. After the phase transition is complete, the $\ps$ fermions have acquired masses
$m_\ps \sim h v_\vf$ and the SU(2)$_G$ gauge bosons have masses $\MG = \frac{1}{2}\gG \: v_\vf$. 
All interactions that follow preserve $B-L$ and $X$.

The $\ps$ fermions now begin sharing the $X$ asymmetry with the 
visible and dark sector particles $f$ and $\x$, and the visible and dark degrees of freedom they couple to, through the Yukawa interaction
\beq
\d \cal{L}_\rm{conn} = -\k \bar{\ps}_L \x f_R + \hc \ ,
\label{eq:Lconn}
\eeq
For reasonable values of $\kappa$,
this interaction keeps the generative sector in equilibrium with the visible and the dark sectors, down to temperatures below the generative phase transition and till about  $T \sim m_\ps/2$, when the $\psi$ inverse decays freeze out.
At that temperature, the abundance of the $\psi$ fermions has become thermally suppressed, and they only carry a small portion of the $X$ asymmetry, having transmitted most of it to relativistic visible and dark sector particles. 
Moreover, the remaining population of $\psi$ fermions will decay, transferring the $X$ asymmetry entirely into equal and opposite $B-L$ asymmetries in the visible and the dark sectors.
If the decay of $\ps$ into $\x$ and $f$ is kinematically possible, it will be completed above the electroweak phase transition provided that 
\beq 
\G_{\ps \to f\x} = \frac{\k^2}{16 \p} m_\ps  \:  \gtrsim  \:  \frac{T\ssub{EW}^2}{M_\rm{P}} \ ,
\label{eq:psi dec rate}
\eeq
which, for $T\ssub{EW} \approx 200\ \rm{GeV}$, leads to the weak constraint
\beq 
\k \gtrsim 4 \cdot 10^{-8} \pare{\frac{2 \TeV}{m_\ps}}^{1/2} \ .
\label{eq:kappa} 
\eeq
If $\ps \to f \x$ is not kinematically allowed, $\psi$ can decay via off-shell $\x$ and/or $f$ particles into lighter particles to which they couple. 
In this model, it is important that the transfer of the $X$ asymmetry to the visible and dark sectors occurs before the electroweak phase transition, because, as we shall see in the next section, the coupling to the visible sector is leptonic, and the action of the electroweak sphalerons is required to process the lepton asymmetry into visible baryonic charge.

The single physical Higgs particle remaining from the $\vf$ multiplet can decay into, for example, electroweak Higgs bosons through a Higgs potential coupling term, and the $G$ gauge bosons decay into the $\ps$ fermions provided their mass is more than twice the mass of the lightest $\ps$ particle.

\subsection{The visible sector}
\label{sec:vis sect}

The vector-like fermions $f$ are SM-gauge-neutral additions to the visible sector, whose role is to transfer the $X$-asymmetry held initially by the generative $\psi$ fermions into a lepton asymmetry in the visible sector, via the interactions
\beq
\d \cal{L}_\rm{v}  = -\yv \: \bar{\ell}_L H f_R - m_f \: \bar{f}_L f_R + \hc  \ .
\label{eq:Lvis} 
\eeq
The lepton asymmetry, if carried by SU(2)$_L$ charged fields, is processed at temperatures above the electroweak phase transition by electroweak sphalerons, and results in a visible baryonic asymmetry.

The fermions $f$ are thus required to have mass greater than the electroweak scale, so that they decouple, decay and transfer all of the $\Lv$ charge to the SM leptons, which couple to the weak interactions, before the electroweak phase transition. They decay via the modes
\beq 
f \rightarrow e_L^- h^+ \ \ \ \\ {\rm and} \ \ \  \ f\rightarrow \n_L h^0 \ ,
\label{eq:f dec} 
\eeq
with rate
\beq 
\G_f = 6 \cdot \frac{\yv^2}{16 \p} m_f    \:  \gtrsim  \:  \frac{T\ssub{EW}^2}{M_\rm{P}} \ ,
\label{eq:f dec rate} 
\eeq
where the factor of six is due to the two decay channels for each of the three families. We therefore require
\beq 
\yv \gtrsim 2 \cdot 10^{-8} \pare{\frac{1 \TeV}{m_f}}^{1/2}  \ .
\label{eq:yv>} 
\eeq

The fermions $f_R$ couple in the same way as right-handed neutrinos to SM fields, but they do not play the same role. In the absence of a neutrino mass generation mechanism, one linear combination of $f_L$ and $\n_L$ remains massless and must be identified with the physical left-handed neutrino prior to neutrino mass generation. Below the electroweak phase transition, the relevant interactions from \eq{eq:Lvis} become: 

\beq
\cal{L} \ \supset \ -m_f \: \bar{f}_L f_R - \frac{\yv v_H}{\sqrt{2}} \: \bar{\n}_L f_R + \hc 
= -\pare{m_f^2 + \frac{\yv^2 v_H^2}{2}}^{1/2} \bar{f}_L' f_R  + \hc \ ,
\eeq
where
\beq
f_L' \equiv \frac{f_L + \th \n_L }{\sqrt{1 + \th^2} } \ ,
\eeq
with $\th^2 = \yv^2 v_H^2/2 m_f^2$ being the mixing parameter between $f$ and $\n$. Accelerator bounds require $\th^2 < 10^{-3}$ for $m_f \gtrsim 1 \GeV$~\cite{Smirnov:2006bu}, which means we must require
\beq 
\yv \leq 0.2 \pare{\frac{m_f}{1 \TeV} } \ .  
\label{eq:yv<} 
\eeq
The lower and upper bounds \eqref{eq:yv>} and \eqref{eq:yv<} allow for a large range of acceptable $\yv$ values, for a given $m_f$.

Through the decays of $f$, the $\BLv$ charge has now been transferred to the SM sector, 
and is reprocessed by sphalerons into a baryon asymmetry,
\beq 
\h(\Bv) = c \: \frac{\h(X)}{2} \approx 5 \times 10^{-10} \ ,
\eeq
where $c \approx 0.3$~\cite{Harvey:1990qw}, and the story in the visible sector is complete.  The final visible matter relic abundance is 
\beq
\W_\rm{VM} \simeq 0.046 = m_p \sqpare{ \h(\Bv) \cdot s} / \r_c = c \: m_p \: \frac{s}{\r_c} \: \frac{\h(X)}{2} \ ,
\label{eq:W VM}
\eeq
where $s$ is the total entropy of the universe today.

\subsection{The dark sector}
\label{sec:dark sect}
The $X$ asymmetry is transferred to the dark sector particles via the scalar $\x$. 
The relation between the visible and dark sector asymmetries given in \eq{eq:Bv-Bd separation} prescribes a relation between the VM and DM relic number densities only if the symmetric thermal population of the DM particles is efficiently annihilated. For this purpose, the DM annihilation cross section has to exceed the canonical value of symmetric thermal WIMP DM, albeit only by a factor of a few~\cite{Graesser:2011wi}. 
However, effective interactions which would annihilate DM into light visible sector particles are typically constrained to be weaker than the weak force, and thus not sufficient for this purpose~\cite{Bai:2010hh,Goodman:2010ku,Buckley:2011kk}. This is, for example, the case with the gauged $B-L$, which is already present in our model.\footnote{\label{foot:ZB} We note that collider constraints on a $Z'_B$ are much weaker~\cite{Williams:2011qb}, and it is possible that such a force could be employed in this and other baryon-symmetric scenarios, and provide efficient annihilation of DM. However, this would require the existence of exotic quarks cancelling the $\rm{[SU(2)}_L]^2 \times \Bv$ anomaly.}

The above considerations compel the existence of a dark gauge force and the annihilation of DM into dark-sector radiation.
We thus extend the dark sector by two Dirac fermions $\ks$ and $\zeta$, which are also charged under the dark gauge group U(1)$_D$.
This force need not be spontaneously broken, and in that sense is a dark-sector analogue of electromagnetism.  
In addition to the U(1)$_D$ gauge force, the dark sector physics is given by
\begin{equation}
\d \cal{L}_\rm{d}   =  - \left( \yd \: \bar{\ks} \x \z + \hc \right) - m_\ks \: \bar{\ks} \ks - m_\z \: \bar{\z} \z - V_\x \ ,
\label{eq:Ldark}
\end{equation}
where $V_\x$ is the scalar potential of $\x$ and we require that $\x$ has a zero VEV. In \eq{eq:Ldark}, we introduced bare masses for the fermions, which will eventually be determined by the requirement of getting the correct DM mass density from the dark asymmetry of \eq{eq:Bv-Bd separation}. It is presumed that the DM mass scale arises from some other physics (e.g Higgs mechanism, confining force), which we do not specify here.

The $\x$ particles decouple, decay and transfer the $X$ asymmetry into the $\ks, \z$ fermions of the dark sector via Eq.~(\ref{eq:Ldark}),
at rate
\beq \G_{\x \to \ks \bar{\z}} = \frac{\yd^2}{16 \p} m_\x  \ .
\eeq
The asymmetry transfer, as well as the DM annihilation and decoupling, have to be completed before matter-radiation equality, so that DM density perturbations can begin to grow. This puts only a very weak constraint on the Yukawa coupling
\beq 
\yd > 10^{-16} \pare{\frac{100 \GeV}{m_\x}}^{1/2} \ . 
\eeq

The three Dirac fermions $\ks_{1,2}$ and $\z$ are stable\footnote{Linear combinations of $\Bd = [ X - (B-L) ]/2$ and $D$ define two dark fermion number symmetries, one carried by the $\ks$ doublet and the other carried by $\zeta$.} and carry all of the antibaryonic asymmetry. 
After they become non-relativistic, their symmetric thermal populations annihilate away via the unbroken U(1)$_D$ force.

We now calculate the relic DM mass density, and compare it with the VM.
The SU(2)$_G$ global symmetry sets $\h(\ks_1) = \h(\ks_2) \equiv \h(\ks)$. 
The conserved the $\Bd$ and $D$ charges are
\bea 
\h(\Bd) &=& \h(\z)                   = \frac{\h(X)}{2} \ , \label{eq:Bd asymm}  \\
\h(D)   &=& \h(\z) + 2 \h(\ks) = 0 \ . \label{eq:D asymm}
\eea
Equations~\eqref{eq:Bd asymm} and \eqref{eq:D asymm} yield
\beq \h(\z) = -2 \h(\ks) = \frac{\h(X)}{2} \ . \label{eq:z, ks asymm} \eeq
The dark matter relic abundance is
\beq
\W_\rm{DM} = \frac{ m_\z |\h(\z)| \cdot s + 2m_\ks |\h(\ks)| \cdot s }{\r_c} = \pare{ m_\z + m_\ks } \frac{s}{\r_c} \frac{\h(X)}{2}  \ .
\label{eq:W DM}
\eeq
Equations \eqref{eq:W VM} and \eqref{eq:W DM} give
\beq
\frac{\W_\rm{DM}}{\W_\rm{VM}} = \frac{m_\ks + m_\z}{m_p} \frac{1}{c} \ ,
\eeq
which predicts that the masses of the DM particles should satisfy\footnote{Equation~\eqref{eq:DM mass} gives, in fact, the mass-to-baryonic-charge ratio of DM in baryon-symmetric models. For the dark sector presented here, the baryonic charge of DM is $q_\rm{DM} = -1$. More generally, the DM mass needed to yield the correct DM abundance in baryon-symmetric models is $m_\rm{DM} \simeq 1.5 \GeV \times |q_\rm{DM}|$ if the $X$ asymmetry is generated and released in the thermal bath above the electroweak phase transition, or  $m_\rm{DM} \simeq 5 \GeV \times |q_\rm{DM}|$ if this happens below the electroweak phase transition. Possible dark-sector sphaleron-like effects could modify these predictions by a factor of a few.}
\beq
m_\ks + m_\z = c \, m_p \: \frac{\W_\rm{DM}}{\W_\rm{VM}} \approx 1.5  \GeV \ .
\label{eq:DM mass}
\eeq
We note that \eq{eq:z, ks asymm}, and therefore the prediction of \eq{eq:DM mass}, would not change if we had shifted the $\Bd$ charges of the $\ks, \z$ fermions by the same amount. This would in fact be equivalent to kinetic mixing of U(1)$_D$ with U(1)$_{B-L}$.

In the above we showed that, in this model, the DM is antibaryonic and consists of the $\bar{\ks}_{1,2},\z$ fermions. Due to the unbroken U(1)$_D$, these particles combine in the late universe into two kinds of $D$-neutral hydrogen-like bound states, $\bar{\ks}_1\z$ and $\bar{\ks}_2\z$ after the decoupling of the $D$-photons.  In this imagined world, the DM today is ``atomic'' as well as asymmetric. Atomic DM entails interactions that are more complicated than those of WIMP DM, and is subject to constraints from cosmological and astrophysical considerations. We discuss pertaining bounds in section~\ref{sec:constraints}.

Despite the $\Bd$ violation necessary to generate the dark baryonic asymmetries, the exact stability of DM is guaranteed by a discrete remnant of $\Bd$ respected by the generative sphalerons. The latter violate $X$ by 4 units: $\delta_g X = |q_X(\psi)| N_f(\psi) = 4$, where $|q_X(\psi)|=2$ is the $X$ charge of the generative fermions and $N_f(\psi) =2$ is the (minimum) number of generative fermionic families. The dark baryonic number $\Bd = [X - (B-L)]/2$ is thus violated by $\delta_g \Bd = \delta_g X /2 = 2$. (Of course this violation is accompanied by $\delta_g \BLv = \delta_g X/2 = 2$ to maintain $\delta_g (B-L)=0$.) This ensures the stability of DM -- even if the generative sphaleron rate were important at zero temperature -- whose decay would require processes inducing $\delta \Bd = 1$.

\subsection{Gauged $B-L$ and its spontaneous breaking}
\label{sec:B-L}

As discussed in section~\ref{sec:symmetrystructure}, baryon-symmetric models rely on the existence of an always conserved particle number, what we called here $B-L$, under which both nucleons and dark matter are charged. In gauge theories, global symmetries of low-energy effective models are expected to be violated at some higher scale, and ultimately by gravity.
The exact conservation of a charge is thus considered technically natural if it can be ascribed to a gauge symmetry.\footnote{Of course, the breaking of a low-energy global symmetry by higher dimensional operators does not necessarily result in production of a net charge.}

If the conservation of $B-L$ is attributed to it being gauged, then it also has to be broken to be consistent with the non-observation of a long-range force. The minimal way to do this is to introduce a scalar $\sigma$ which has a nonzero $B-L$ but which transforms trivially under all the other symmetries, and require it to acquire a nonzero VEV. Here, we require that the breaking of the gauged $B-L$ leaves a global symmetry unbroken, under which all the particles of Table~\ref{tab:charges} and the SM carry their usual $B-L$ gauge charges. 
To achieve this, we disallow Yukawa interactions of the following kind:  
$\s^n \cal{O}_{B-L}$, with $n$ any positive integer and $\cal{O}_{B-L}$ being a spin-0 monomial consisting of SM fields and the fields of Table~\ref{tab:charges}, which carries $B-L \ne 0$ and is invariant under all other symmetries (including $X$). Such a Yukawa coupling would break the global $B-L$ we wish to retain upon $\s$ developing a VEV.

This structure, which allows for the breaking of a local U(1) to a global remnant, can be also described as follows: the gauge symmetry encompasses two global U(1) symmetries, only one of which breaks spontaneously and gives mass to the gauge boson. In our example we can write 
\beq
(B-L)_\rm{gauged}  = (B-L)_\rm{global} + B' \ .
\eeq 
The SM and the particles of Table~\ref{tab:charges} transform under $(B-L)_\rm{global} = \BLv - \Bd$ as described in the text, and have $B'=0$. The field $\s$ carries $B' \ne 0$ and $(B-L)_\rm{global}=0$, obtains a VEV and breaks the gauge symmetry. Note that $X = \BLv + \Bd$, and the charge of $\s$ under $X$ is $q_{_X} (\s) = 0$. The phase transition associated with the $B-L$ breaking will not alter the cosmological asymmetry because, by construction, $\sigma$ does not carry $X$ or $(B-L)_\rm{global}$ charge.

We should now identify what the $B-L$ charge of $\s$, $q_{_{B-L}} (\s)$, should or should not be, such that the above symmetry structure (and thus symmetry breaking pattern) arises accidentally, as a result of the $B-L$ gauge charges of the fields, rather than from the imposition of a global $B'$ symmetry.
The accidental global $B'$ certainly occurs for interactions of all orders if $\s$ carries an irrational $B-L$ charge. 
If, on the other hand, $\s$ carries integer or fractional $B-L$ charge, then the global $B'$ symmetry emerges at a finite order of interactions.
For example, given the particle content of our model, the  lowest-order $\cal{O}_{B-L}$ operators (as defined above) are the dimension-4 monomials $\bar{\ps}_L^{} \x f_L^c + \rm{h.c}$.
Thus, no renormalisable Yukawa interaction violates the global $B'$ for any value of $q_{_{B-L}} (\s)$. In addition, if $q_{_{B-L}} (\s) \ne \pm  2$, the Yukawa interactions $\s (\bar{\ps}_L^{} \x f_L^c)$ and $\s^* (\bar{\ps}_L^{} \x f_L^c)$ are forbidden by gauge invariance, and the $B'$ symmetry is preserved up to interactions of at least order 5. Additional conditions arise if we require $B'$ to be respected by higher-order operators.\footnote{
The spin-0 monomials 
$ \tilde{\vf} \, \x,     
\ \overline{\ps}_R f_L ,         
\ \overline{\ps_R^c} f_R^{},     
\ \overline{\ps_R^c} \n_R^{}, 
\ \overline{f_L^c}   f_L^{},    
\ \overline{f_R^c}   f_R^{},    
\ \overline{\n_R^c} \n_R^{},    
\ \overline{\n_R^c}  f_R^{}$  
have lower dimensionality 
than $\bar{\ps}_L^{} \x f_L^c$
and carry no gauge charge other than $B-L \ne 0$. However, they also carry $X$ charge. Since we have chosen to impose a global $U(1)_X$ symmetry (see section~\ref{sec:gen sect} and footnote~\ref{foot:Xsym}), under which $\s$ does not transform, we need not consider these monomials here. However, considering them would only imply $q_{_{B-L}} (\s) \ne  \pm 1/2, \pm  1, \pm 2$.}
The order at which $B'$ emerges as a global symmetry of the unbroken theory, is, of course, the order at which $(B-L)_\rm{global}$ is retained after $\s$ acquires a VEV and breaks the gauged $B-L$.

When the scalar $\s$ acquires a nonzero VEV $v_\s$, the gauge boson $\ZBL$ acquires a mass $\MBL = \gBL v_\s/2 $, where $\gBL$ is the gauge coupling constant. 
Collider experiments constrain $\ZBL$ parameters so that the lower bound on the mass is $\sim 500 \GeV$ for $\aBL =\gBL^2/4\p \sim 10^{-2}$~\cite{Williams:2011qb}.  
We discuss the scale of $B-L$ breaking in the next section, in association with the generative and the electroweak phase transitions.

\subsection{The scalar degrees of freedom}
\label{sec:scalars}

Our model contains several scalar degrees of freedom: $\vf, \s, \x$ and the SM Higgs $H$.
All of them are subject to their scalar potentials, which at tree-level and zero temperature are
\begin{alignat}{10}
& V_\rm{G}  &     &= \ &     &-&  &\m_\vf^2 |\vf|^2 &     & \: + \: &    & \l_\vf |\vf|^4 & , \label{eq:VG} \\
& V_\s      &     &= \ &     &-&  &\m_\s^2  |\s|^2  &     & \: + \: &    & \l_\s  |\s|^4  & , \label{eq:Vsigma} \\
& V_H       &     &= \ &     &-&  &\m_H^2   |H|^2   &     & \: + \: &    & \l_H   |H|^4   & , \label{eq:VH} \\
& V_\x      &     &= \ &     & &  &\m_\x^2  |\x|^2  &     & \: + \: &    & \l_\x  |\x|^4  &   \label{eq:Vchi} \ .
\end{alignat}
We want the fields $\vf, \s, H$ to acquire VEVs and break the generative, $B-L$ and electroweak symmetries, while $\x$ should not acquire a VEV (thus leaving $\Bd$ unbroken).
 We therefore require  $\m_\vf^2, \: \m_\s^2, \: \m_H^2, \:  \m_\x^2 >0$.
The full scalar potential of the theory also contains terms that couple the various scalar particles together
\beq
\bal{6}
V_\rm{mix}(\vf, \s, H , \x) 
=& \ \l_{\vf H} |\vf|^2 |H|^2 \ &+& \ \l_{\vf \x} |\vf|^2 |\x|^2 \ &+& \ \l_{\vf \s} |\vf|^2 |\s|^2 \\
+& \ \l_{\s H}  |\s|^2  |H|^2 \ &+& \ \l_{\s  \x} |\s|^2  |\x|^2 \ &+& \ \l_{H \x}  |H|^2   |\x|^2 \ .
\eal
\label{eq:Vmix}
\eeq
We collectively refer to the couplings of \eq{eq:Vmix} as $\l_\rm{mix}$, and to the quartic self-couplings of \eqs{eq:VG} to \eqref{eq:Vchi} as $\l_\rm{self}$.

After the $\vf, \s, H$ fields acquire VEVs, the mixed quartic couplings generate mass contributions for all the scalars, $m_\rm{ind}^2 \sim \l_\rm{mix} v^2$, and induce mixing among the $\vf, \s, H$ fields.
The requirement that $\x$ does not acquire a VEV becomes $\m_\x^2 + \l_{\vf \x} v_\vf^2 + \l_{\s \x} v_\s^2 + \l_{H \x} v_H^2 >0$.
We may discern two cases, depending on the relative magnitudes of $\l_\rm{mix}$ and $\l_\rm{self}$:
\benu[(i)]

\item Small mixing, $\l_\rm{mix} \ll \l_\rm{self}$.  

This regime can arise naturally from an underlying supersymmetric dynamics for mixing between fields with no gauge interactions in common; in particular, zero mixing follows for the $D$-term contributions to the potential.

For this case, the mass contributions induced by symmetry breaking are much smaller than the mass-parameters in the Lagrangian, $m_\rm{ind}^2 \ll \m^2$, and the mixing among the scalar degrees of freedom that acquire VEVs is 
\beq
\th_{ij} \sim \frac{\l_\rm{mix} }{ \l_\rm{self} } \frac{\min (\m_i,\m_j)}{\max (\m_i,\m_j)} \ll 1 \ .
\eeq
The phase transitions which break the various symmetries proceed independently, and a hierarchy of scales can be accommodated.  We require
\beq
\MG > M_{_\rm{EW}}, \  m_\x \ ,
\label{eq:M hier, no mix}
\eeq
such that the $X$ asymmetry generation occurs above the electroweak phase transition, allowing for leptonic couplings to be employed for the transfer of the asymmetry from the generative sector to the relic visible sector particles, as in the example presented here\footnote{It is, of course, possible to implement our scenario using baryonic couplings between the generative and the visible sectors.}. 
While the $\x$ field does not strictly need to be lighter than the generative sector particles, this would allow for unsuppressed decay of the $\ps$ fermions. However, the decay of $\ps$ can also proceed via an off-shell $\x$ at a sufficient rate, if this is kinematically necessary.

For $\l_\rm{mix} \lesssim 10^{-6}$, the mixed quartic couplings cannot keep the various sectors in equilibrium, but they may contribute to the scalar decay width significantly.

\item Significant mixing, $\l_\rm{mix} \sim \l_\rm{self}$.

This is the most generic possibility in non-supersymmetric models and if the scalars are fundamental particles.

The induced mass contributions from symmetry breaking are large, and as a result they balance the various scales involved. Only small hierarchies among scales can be sustained, and we assume that
\beq
\MG \sim \MBL \gtrsim M_{_\rm{EW}} \sim m_\x \ .
\label{eq:M hier}
\eeq
The mixing between the scalar degrees of freedom which acquire VEVs is significant. This is of particular importance for collider physics, as it relaxes the limits on the SM Higgs mass, both because of suppression  in the Higgs production cross-section in colliders, and suppression in the Higgs decay branching fraction into SM states.

The couplings of \eq{eq:Vmix} contribute to the thermalisation and decay processes among the various sectors, which can be important for the cosmological viability of the model, as we now discuss.
\eenu

\section{Cosmological and astrophysical constraints}
\label{sec:constraints}

We now turn to examining some aspects of the cosmology of our model arising from the existence of a dark sector with several stable particle species and internal dynamics.

The visible and the dark sectors are kept in equilibrium with each other in the early universe through their couplings~(\ref{eq:Lconn}) to the generative sector via $\ZBL$ exchange, and through the scalar potential term $\l_{H \x} |\x|^2 |H|^2$ of \eq{eq:Vmix}. 
If the dark U(1)$_D$ mixes kinetically with hypercharge, via the term $-(\e/2) F_{_Y}^{\m\n} F_{_D \: \m\n}^{}$, the scattering processes $e^- \g \leftrightarrow e^- \gD$ can equilibrate the two sectors down to $T \sim m_e$. We require $\e \lesssim 10^{-8}$ to preclude this possibility.
Depending on whether the scalar coupling $\l_{H \x}$ is small or large, as discussed in the previous section, the dark and the visible sectors decouple at a temperature
\beq 
\rm{(i)}    \   \Tdec \sim \min \sqpare{ \MG, \MBL }, 
\qquad \rm{or} \qquad 
\rm{(ii)}   \   \Tdec \sim M_{_\rm{EW}} \ .
\label{eq:Tdec} 
\eeq

Below $\Tdec$ the visible and dark sectors are in equilibrium within themselves, but have different temperatures $\Tv$ and $\Td$, respectively.
Entropy conservation relates their temperatures according to
\beq 
\frac{\gv \Tv^3}{\gd \Td^3} = \frac{\gv(\Tdec)}{\gd(\Tdec)} \equiv r  \ ,
\label{eq:Trel} 
\eeq
where $\gv (\gd)$ is the number of relativistic degrees of freedom (d.o.f.) in the visible (dark) sectors.
As the universe evolves after visible-dark-sector decoupling, the ratio of temperatures changes from unity as mass thresholds are crossed, 
particles become cold and cause the $g$'s to decrease.   
We use \eq{eq:Trel} to express the entropy and the energy density of the universe, and other related quantities, at any point after $\Tdec$, in terms of either $\Tv$ or $\Td$.

\subsection{BBN and WMAP}

We must ensure that the energy density due to relativistic dark-sector species does not spoil big bang nucleosynthesis (BBN) and WMAP measurements.
The observational limit on the additional relativistic energy density at temperatures $\lesssim 1 \MeV$ is customarily expressed through an ``effective number of extra neutrino species'', $\d N_\rm{eff}$. Recent BBN analysis~\cite{Mangano:2011ar} suggests $\d N_\rm{eff} \leqslant 1$ at $95\%$~C.L., while the current best fit from WMAP is $\d N_\rm{eff} = 1.34^{+0.86}_{-0.88}$ at $68\%$~C.L.~\cite{Komatsu:2010fb}. For our purposes, we use $\d N_\rm{eff} \leqslant 1$.

The relativistic energy density carried by the dark sector depends both on the number of relativistic d.o.f. and on the temperature of the dark sector. The dark-sector temperature can be lower than that of the visible sector if the fraction of relativistic d.o.f. which decouple and release their entropy below $\Tdec$ is larger in the visible sector than in the dark sector, as exhibited by \eq{eq:Trel}.

For our model, the dark-sector relativistic d.o.f. at temperatures $\lesssim 1$~MeV are the two massless U(1)$_D$ gauge boson polarisation states, $\gdN = 2$, while for the visible sector $\gvN = 3.91$. This sets a bound on the dark relativistic d.o.f at $\Tdec$
\beq
\gdd \leqslant 18.6 
\pare{ \frac{\gvd}{110.25} }
\ .
\label{eq:gdd upper}
\eeq

If the visible-dark-sector decoupling happens above the electroweak phase transition, when all of the SM d.o.f and the $f$ fermions are still relativistic, then \eq{eq:gdd upper} yields $\gdd \leqslant 18.6 $. The dark sector of Table~\ref{tab:charges} has a total $\gd$ of $16.5$, so it satisfies the bound. 

If the mixing couplings in the scalar potential of \eq{eq:Vmix} are large, the visible-dark-sector decoupling happens below the electroweak phase transition, when the SM Higgs, the top quark and the $f$ fermions have decoupled and $\gvd = 92.75$. Because of the large scalar mixing coupling, $\x$ also has an electroweak-scale mass and is decoupled, leaving $\gdd = 12.5$. The bound of \eq{eq:gdd upper} now reads $\gdd \leqslant 15.6 $, and is again satisfied.

Any additional heavy particles in the visible sector would allow for additional dark degrees of freedom.

\subsection{Dark recombination and residual ionisation}

The dark matter of our model consists of the $\bar{\ks}_{1,2}, \z$ fermions which carry U(1)$_D$ charge. They form Hydrogen-like states $\bar{\ks}_1 \z$ and $\bar{\ks}_2 \z$, thus screening the long-range force among the DM particles. We have to ensure that the dark recombination, $\bar{\ks} + \z   \to   \HD + \gD$, occurs before the matter-radiation equality, so that the $D$-neutral DM states can start clustering and density perturbations can begin to grow. To estimate the recombination time and the residual ionisation fraction, we employ a simple thermal equilibrium and freeze-out calculation.\footnote{For an overview of constraints on atomic DM, see Ref.~\cite{Kaplan:2009de} and references therein.}

The dark baryon density and ionisation fraction are
\begin{gather}
n(\Bd) = n(\z) + n(H_D) \ ,     \label{eq:nBd} \\
X_\z \equiv \frac{n(\z)}{n(\Bd)} = \frac{n(\ks)}{n(\Bd)} \equiv X_\ks  \ , \label{eq:Xz}
\end{gather}
where $n(\ks) = n(\ks_1) + n(\ks_2)$ and  similarly for the dark Hydrogen states.
The dark baryon asymmetry is fixed by the observed visible baryonic asymmetry
\beq
\h(\Bd) = \frac{n(\Bd)}{s} = \frac{1}{c} \h(\Bv) \approx 10^{-9} \ .  \label{eq:etaBd}
\eeq
The (ground-state) binding energy of the dark Hydrogen is
\beq 
E_b  = \frac{\aD^2 \mD}{2} = \frac{\aD^2}{2} \frac{m_\ks m_\z}{m_\ks + m_\z} \ ,  \label{eq:B} 
\eeq
where $\aD$ is the U(1)$_D$ fine-structure constant and $\mD$ is the $\z,\bar{\ks}$ reduced mass.
In what follows, we adopt the ratio
\beq
x \equiv \frac{E_b}{\Td}  \label{eq:x}  
\eeq
as our time parameter (with $\Td$ being the dark-sector temperature).
As long as the recombination reactions $\bar{\ks} + \z   \leftrightarrow   \HD + \gD$ are in equilibrium, the Saha equation gives the ionisation fraction
\beq
X_\z^\rm{EQ} (x)  =  \cal{X} \sqpare{\aD, \h(\Bd)}  \ x^{3/4} \: e^{-x/2} \ ,  
\label{eq:XzEQ}
\eeq
where we have taken the limit  $X_\z \ll 1$, and defined for convenience
\beq \cal{X} \sqpare{\aD, \h(\Bd)} \equiv \pare{ \frac{45  (1+r)}{2 \p^{7/2} \gd} }^{1/2}  \aD^{-3/2} \: \h(\Bd)^{-1/2} \ , \label{eq:calX} \eeq
with $r$ defined in \eq{eq:Trel}.
Recombination is defined as the moment when $X_\z^\rm{EQ}(x_\rm{rec}) = 0.1$, which yields
\beq 
x_\rm{rec} \simeq 34.5  + \ln \sqpare{ \pare{\frac{0.3}{\aD} }^3 \pare{ \frac{10^{-9}}{\h(\Bd)} } } \ .  
\label{eq:x rec sol} 
\eeq
Dark recombination has to happen before matter-radiation equality, at visible sector temperature $\Tv \approx 1 \eV$, which is ensured for sufficiently large dark-Hydrogen binding energy or reduced mass
\beq
\mD \gtrsim 0.5 \keV \pare{\frac{0.3}{\aD}}^2 \ .
\label{eq:mu min}
\eeq
This lower limit on the $\z, \ks$ reduced mass $\mD$ is much smaller than the sum of masses $m_\z + m_\ks \approx 1.5 \GeV$ (see \eq{eq:DM mass}), and implies that the lightest of the $\ks, \z$ fermions should be at least as heavy as $\mD$.

In the above analysis we assumed that the recombination reactions $\bar{\ks} + \z   \leftrightarrow   \HD + \gD$ stay in equilibrium at least till the ionisation fraction drops to about 10\%. We now discuss the constraints under which this holds, and what is the residual dark ionisation, after the freeze-out of the recombination reactions.

The thermally averaged cross-section of the recombination reactions is~\cite{Peebles:1968ja}
\beq \ang{ \s v } = \tilde{c} \frac{64 \p}{\sqrt{27 \p}} \frac{\aD^2}{\mD^2} x^{1/2} \ln x \ , \label{eq:sigma recomb} \eeq
with $\tilde{c} \simeq 0.448$. Note that recombination proceeds primarily into the first excited state of the dark Hydrogen, which then decays into the ground state. 
Freeze-out occurs when $H = \ang{ \s v } \: n(\z)$. Assuming radiation-domination, which is required for the pre-dark-recombination era, this condition yields the time at which the ionisation fraction freezes out
\beq 
x_f - \frac{1}{2} \ln x_f - 2 \ln (\ln x_f) =  \cal{F} \sqpare{ \aD, \h(\Bd), \mD } \ ,  \label{eq:xf eq}
\eeq
where we have defined
\bea 
\cal{F} \sqpare{\aD, \h(\Bd), \mD} 
&\equiv& 
\ln \sqpare{ \frac{2^9 \: \tilde{c}^2}{3^3 \p^{3/2}} \: \frac{(1+r)^3}{1+\tilde{r}} } + 
\ln \sqpare{ \h(\Bd) \aD^5 \pare{\frac{M_P}{\mD}}^2 } \ ,
\eea
with $\tilde{r} = r^{4/3} \pare{\gd/\gv}^{1/3}$.
Thus, the recombination reactions freeze out at
\beq 
x_f \simeq 72 + \ln \sqpare{ \pare{\frac{\h(\Bd)}{10^{-9}}} \pare{ \frac{\aD}{0.3} }^5  \pare{\frac{0.4 \GeV}{\mD}}^2 } \ .
\label{xf}
\eeq
The residual ionisation fraction $X_\z^\rm{res} = X_\z^\rm{EQ} (x_f)$ is
\beq
X_\z^\rm{res} \approx 10^{-9} \pare{ \frac{10^{-9}}{\h(\Bd)} } \pare{ \frac{0.3}{\aD} }^4 \pare{ \frac{\mD}{0.4 \GeV} } \ .
\label{eq:Xz res}
\eeq
Requiring $X_\z^\rm{res} < 0.1$ (the recombination value) yields
\beq
\aD \geqslant 3 \times 10^{-3} \pare{ \frac{10^{-9}}{\h(\Bd)} }^\frac{1}{4}  \pare{ \frac{\mD}{0.4 \GeV} }^\frac{1}{4} \ .
\label{eq:alpha min ion}
\eeq

\medskip

Equations~\eqref{eq:mu min} and \eqref{eq:alpha min ion} summarise the mild constraints that arise from requiring that long-range self interactions of the DM are sufficiently screened in the early stages of gravitational collapse to allow for clustering and the formation of inhomogeneities. 
These estimations are consistent with more detailed numerical methods~\cite{Ma:1995ey}.

\subsection{Dark-matter self interaction inside clusters}
If dark matter is atomic, then atom-atom collisions can be important inside galaxies and clusters. 
Analyses of the Bullet Cluster bound the DM self-interaction~\cite{Markevitch:2003at,Randall:2007ph} by
\beq
\frac{\sigma}{m_\rm{DM}}  < 1 \cm^2/ \gr  \ .
\label{eq:bullet con}
\eeq
The low-energy atom-atom cross section can be estimated as~\cite{Kaplan:2009de}
\beq
\s \approx 4 \p k^2 a_0^2 \ ,
\eeq
where $a_0 = \pare{\aD \: \mD}^{-1}$ is the Bohr radius, and $3 \leqslant k \leqslant 10$.
Thus, the Bullet Cluster constraint implies
\beq
\aD \geqslant 0.3 \pare{ \frac{k}{3} }  \pare{ \frac{0.4 \GeV}{\mD} } \pare{ \frac{1.5 \GeV}{m_\rm{DM}} }^\frac{1}{2} \ .
\label{eq:alpha min bullet}
\eeq
In our model $m_\rm{DM} = m_\z + m_\ks \approx 1.5 \GeV$ (see \eq{eq:DM mass}), and for $m_\z \approx m_\ks$ the reduced mass is $\mD \approx 0.4 \GeV$. 
We note that $\aD \sim 0.3$ is well within the perturbative range.

In the above, we ignored the possibility of DM having a small ionised component, a remnant of the freeze-out of the recombination reactions, which would increase the self interactions of DM inside clusters. However, the constraint of \eq{eq:alpha min bullet} ensures that the residual ionisation fraction, estimated by \eq{eq:Xz res}, is tiny and completely negligible. 

Equation \eqref{eq:alpha min bullet} is the strongest constraint pertaining to the self-interactions of dark matter in the model presented here.
Quite importantly, it automatically precludes the possibility of efficient cooling of DM halos due to DM collisions, excitations and de-excitations via the emission of dark photons. This is because the bound of \eq{eq:alpha min bullet} sets a significant lower limit on even the smallest energy splittings of the dark atoms: the hyperfine structure of Hydrogen-like atoms is 
$\d E_\rm{hf}/ E_b \sim \aD^2 \sqpare{ \min(m_\z, m_\ks) / \max(m_\z, m_\ks) }$, which for $\aD \sim 0.1 - 1$ and $m_{\z,\ks} \sim 1 \GeV$ is $\d E_\rm{hf} \sim 10 \keV - 100 \MeV$. Such energy splittings are too large to be excited via atomic collisions inside the DM protohalos, and thus cooling via emission of dark photons does not follow.

\section{Signatures}
\label{sec:signatures}

\subsection{Collider signatures}

A massive $\ZBL$ gauge boson arises naturally in baryon-symmetric models. The scale of $B-L$ breaking need not be, and perhaps cannot be, very different from the electroweak scale, as discussed in section~\ref{sec:scalars}. In this case, the $B-L$ interaction can be probed at the LHC. The Higgs and $\ZBL$ phenomenology at the TeV scale in 
$B-L$ extensions of the Standard Model has been explored in Refs.~\cite{Emam:2007dy,Basso:2009hf,Basso:2008iv,Basso:2010pe,Basso:2010yz,Coutinho:2011xb}.
In models in which the DM is antibaryonic, the  $\ZBL$ is common to both the visible and the dark sectors, and this modifies its phenomenology in interesting ways. 
If produced in colliders, it will have a significant decay width into dark-sector particles, which can be measured at the LHC~\cite{Emam:2007dy,Basso:2009hf,Basso:2008iv,Basso:2010pe,Basso:2010yz,Coutinho:2011xb,
Leike:1998wr,Rizzo:2006nw,Langacker:2008yv,Petriello:2008zr,Coriano:2008wf,Petriello:2008pu,Gershtein:2008bf,Langacker:2009im} 
providing an important probe of the dark sector.

The class of models we have considered entails an enhanced Higgs sector, which provides promising signatures at the LHC. The various Higgs degrees of freedom mass mix, and thus share production and decay channels. The coupling of the electroweak-breaking Higgs to the $(B-L)$-breaking Higgs provides another probe of the baryonic dark sector. 
While this is generic in baryon-symmetric models, the mechanism presented here also predicts the existence of a third Higgs particle which breaks the generative symmetry.
The generative sector connects the visible and the dark sectors, and if a Higgs state with a significant generative-sector component is produced at the LHC, it will decay into multilepton/multijet final states and missing energy. For example, for the particular implementation of the mechanism we presented here, the decay chain of interest would be
\beq
\vf \to \ps + \bar{\ps}, \qquad
\bal{10}
\ps       &\to \x   &\: \rm{( \to missing \ energy)} &\: + \: f,       \qquad  & f        &\to \ell^- W^+ \ \rm{or} \ \n Z \\
\bar{\ps} &\to \x^* &\: \rm{( \to missing \ energy)} &\: + \: \bar{f}, \qquad  & \bar{f}  &\to \ell^+ W^- \ \rm{or} \ \bar{\n} Z \ ,
\eal
\label{eq:colliders}
\eeq
where the intermediate $\ps$, $f$ and $\x$ particles can be off-shell.  The SM gauge bosons can be identified either by their leptonic decays, resulting in high lepton multiplicity, or by their hadronic decays, resulting in high jet multiplicity and allowing the missing energy to be attributed completely to the dark sector. If the generative sector coupling to the visible sector is baryonic rather than leptonic (e.g.\ via the neutron portal $\overline{u_R^c} d_R \overline{d_R^c} f_R$), then the multiplicity of jets in the final state increases. Such a signature has low SM background. It also has the potential to be distinguished from other beyond SM processes (e.g. susy) using the topology of the process and the kinematics reflecting light final dark-sector particles.

We emphasise that such processes are very different from the Higgs decaying part of the time invisibly, purely into missing energy (dark-sector particles). The latter possibility is often encountered in extensions of the SM. However, in the present model the Higgs is decaying into visible and dark-sector particles \emph{simultaneously}, which is indicative of the symmetry structure of baryon-symmetric models.

Because of their couplings to the dark sector, the collider bounds on the $\ZBL$ and the Higgs bosons are relaxed. This is due to suppressed decay branching ratios into SM particles, as well as suppressed production cross-sections in the case of the Higgs. A potential discovery of a Higgs state in the SM-excluded mass range would signal the existence of additional scalar degrees of freedom.

\subsection{Dark-matter direct detection}

For the minimal dark sector constructed here, dark matter can be detected via $\ZBL$ exchange. 
The spin-independent scattering cross section per nucleon can be as high as\footnote{The direct-detection cross section can be larger if DM scattering occurs via exchange of $Z_{_B}'$ rather than $\ZBL$, as discussed in footnote \ref{foot:ZB}.}

\beq
\s_{_{B-L}}^\rm{SI}   \!\!  \approx  
10^{-44} \cm^2  
\pare{ \frac{\gBL}{0.1} }^4 \!
\pare{ \frac{0.7 \TeV}{\MBL} }^4 \ ,
\label{eq:sigmaDD_B-L}
\eeq
where we have used $m_\rm{DM} = 1.5 \GeV$. 
For such low DM masses there are no constraints from XENON100~\cite{Aprile:2011hi} or CDMS II~\cite{Ahmed:2010wy}.

An additional possibility for direct detection arises if U(1)$_D$ is broken and mixes kinetically with the photon. The cross section can be
\beq
\s_{_D}^\rm{SI}    \approx   \pare{10^{-37} \cm^2} 
\pare{ \frac{\e}{10^{-4}} }^2      
\pare{ \frac{g_{_D}}{0.1} }^2     
\pare{ \frac{100 \MeV}{M_{_D}} }^4 .
\label{eq:sigmaDD_dark}
\eeq
This could account for the regions favoured by DAMA and CoGeNT~\cite{Savage:2008er,Aalseth:2010vx} if the DM mass is somewhat higher than what is predicted here. This is possible in a richer dark sector in which the DM particles carry a larger baryonic charge (making their number density smaller and their mass larger). 
We note that a broken U(1)$_D$ provides a suitable dark-sector scenario for our mechanism, provided that the corresponding gauge boson is sufficiently light, $M_{_D} < m_{\z} \sim \GeV$, such that the DM particles can still annihilate into dark photons. Efficient annihilation of the symmetric part of DM requires only $g_{_D} \gtrsim 0.05$. 
For $M_{_D} \sim 100 \MeV$, the kinetic mixing between U(1)$_D$ and the photon can be $\e \sim 10^{-4}$ (for constraints on $\e$, $M_{_D}$ see~\cite{Bjorken:2009mm,Dent:2012mx}). In this case, the dark gauge bosons $Z_{_D}'$ decay as soon as they decouple, via the kinetic mixing into $e^\pm$ pairs, leaving no dark-sector relativistic energy density present at BBN. The breaking of U(1)$_D$ can destabilise one of the dark-sector particles ($\ks$ or $\z$, depending on the mass hierarchy), while $\Bd$ still ensures the stability of the other, which alone constitutes the DM state. There is no long-range force among the DM particles, and the astrophysical constraints of section \ref{sec:constraints} are not applicable. 
The kinetic mixing between the dark force and the photon can be probed in current fixed target experiments~\cite{Bjorken:2009mm,Essig:2010xa}.

\section{Conclusions}

The similar mass densities of the visible and the dark matter of the universe suggest a common origin for both. This can be realised in the context of a baryon-symmetric universe, which entails a dark-sector asymmetry that exactly cancels the visible-sector asymmetry.
While this approach is primarily phenomenological, such a construction is also economical in that the new physics introduced to generate the observed baryonic asymmetry of our universe is also responsible for the dark-matter relic abundance. The dark sector itself need not involve more than few degrees of freedom, although a greater complexity is \emph{a priori} possible, and surely not unparalleled given the complexity of the visible sector.

In this work, we have shown that a baryon-symmetric universe can be achieved through a first-order phase transition associated with the spontaneous breakdown of a new gauge force called the ``generative'' interaction. The possibility analysed here adds to other mechanisms suggested for producing a baryon-symmetric universe~\cite{Dodelson:1989ii,Dodelson:1989cq,Dodelson:1990ge,Kuzmin:1996he,Kitano:2004sv,Kitano:2005ge,Gu:2007cw,Gu:2009yy,An:2009vq,Davoudiasl:2010am,Gu:2010ft,Heckman:2011sw,Oaknin:2003uv,Farrar:2005zd,Bell:2011tn,Cheung:2011if,vonHarling:2012yn,MarchRussell:2011fi,Kamada:2012ht}.
We have presented a specific and minimal dark sector to demonstrate that a viable model of this type exists. 
This dark-sector example features an unbroken Abelian gauge interaction -- a dark-sector electromagnetism -- to annihilate the symmetric part of the dark plasma, and which leads in the later universe to ``atomic'' dark matter.  The stable dark matter fermions form two-body bound states because of the Coulomb interaction of the dark electromagnetism.  We have then derived constraints on this dark sector from big bang nucleosynthesis, structure formation and the Bullet Cluster, and have specified the viable parameter space.

A major phenomenological benefit of baryon-symmetric models, including the one being described here, is the natural appearance of a $Z'$ with a substantial invisible width into dark matter.  The discovery of such a particle at the LHC would be an extraordinary breakthrough in understanding dark matter.  At low energies, the contact interaction provided by $Z'$ exchange can also produce a signal in direct dark matter detection experiments.

The specific realisation developed here is best probed in colliders through the scalar sector.  In addition to the standard electroweak Higgs boson, there is also a scalar that breaks U(1)$_{B-L}$ and another multiplet that breaks the generative symmetry and produces the first-order phase transition that is at the heart of the dynamics.  Naturalness arguments suggest that the generative scalar should mix substantially with the standard Higgs particle.  Such a mass eigenstate can be produced at the LHC through the standard Higgs admixture, but decay through the generative component.  This leads to low-background signatures of multilepton final states or a lepton pair and four jets, plus missing energy carried off by dark sector particles.

We note that the complete explanation of the similarity of the visible and dark mass densities requires a theory of DM mass. In baryon-symmetric models the DM mass is predicted to be in the GeV range. The DM direct detection signals seen by DAMA, CoGeNT and CRESST, albeit in need of confirmation, provide evidence for this mass scale and reinforce the case for similar visible and dark number densities.

\acknowledgments 
KP thanks Benedict von Harling, Tony Limosani and Martin White for helpful discussions, and RRV thanks Jason Kumar. KP and RRV were supported, in part, by the Australian Research Council. The work of MT is supported in part by NASA ATP grant NNX08AH27G, Department of Energy grant DE-FG05-95ER40893-A020 and by the Fay R. and Eugene L. Langberg chair. RRV thanks MT for hospitality at the University of Pennsylvania while part of this work was performed.

\bibliographystyle{JHEP}
\bibliography{Bibliography.bib}

\end{document}